\documentclass[journal]{vgtc}                

\usepackage{mathptmx}
\usepackage{graphicx}
\usepackage{times}
\usepackage{comment}
\usepackage{enumitem}
\usepackage{amsmath}
\usepackage[normalem]{ulem}
\usepackage{xcolor}
\usepackage{makecell}
\usepackage[bookmarks,backref=true,linkcolor=black]{hyperref} 
\hypersetup{
  pdfauthor = {},
  pdftitle = {},
  pdfsubject = {},
  pdfkeywords = {},
  colorlinks=true,
  linkcolor= black,
  citecolor= black,
  pageanchor=true,
  urlcolor = black,
  plainpages = false,
  linktocpage
}
\usepackage{url}

\usepackage{breakurl}

\onlineid{0}

\vgtccategory{Research}

\vgtcinsertpkg

\title{Communicating Visualizations without Visuals: \\ Investigation of Visualization Alternative Text \\ for People with Visual Impairments}

\author{Crescentia Jung, Shubham Mehta, Atharva Kulkarni, Yuhang Zhao, and Yea-Seul Kim}
\authorfooter{
 \item
 Crescentia Jung is with the University of Wisconsin-Madison. E-mail: csjung@wisc.edu.
 \item
 Shubham Mehta is with the University of Wisconsin-Madison. E-mail: smehta23@wisc.edu.
 \item
 Atharva Kulkarni is with the University of Wisconsin-Madison. E-mail: akulkarni23@wisc.edu.
 \item
 Yuhang Zhao is with the University of Wisconsin-Madison. E-mail: yuhang.zhao@cs.wisc.edu.
 \item
 Yea-Seul Kim is with the University of Wisconsin-Madison. E-mail: yeaseul.kim@cs.wisc.edu.
}
\shortauthortitle{}

\abstract{ Alternative text is critical in communicating graphics to people who are blind or have low vision. Especially for graphics that contain rich information, such as visualizations, poorly written or an absence of alternative texts can worsen the information access inequality for people with visual impairments. In this work, we consolidate existing guidelines and survey current practices to inspect to what extent current practices and recommendations are aligned. Then, to gain more insight into what people want in visualization alternative texts, we interviewed 22 people with visual impairments regarding their experience with visualizations and their information needs in alternative texts. The study findings suggest that participants actively try to construct an image of visualizations in their head while listening to alternative texts and wish to carry out visualization tasks (e.g., retrieve specific values) as sighted viewers would. The study also provides ample support for the need to reference the underlying data instead of visual elements to reduce users' cognitive burden. Informed by the study, we provide a set of recommendations to compose an informative alternative text.  } 

\keywords{accessible visualization, assistive technologies, alternative text for graphics}

\CCScatlist{ 
 \CCScat{K.6.1}{Management of Computing and Information Systems}%
{Project and People Management}{Life Cycle};
 \CCScat{K.7.m}{The Computing Profession}{Miscellaneous}{Ethics}
}

\begin{document}



\maketitle

\section{Introduction}
Visualizations are a powerful medium to summarize complex data effectively, as they offload the cognitive burden of a reader by leveraging visual perception~\cite{card1999}. The visualization community has devised theories and principles to design effective visualizations so that readers can extract insights without much mental effort. As a result, visualizations are pervasive across the media, scientific communities, and government agencies to present data. In addition to efficient data communication, visualizations also provide credibility to the content presented in articles~\cite{kong2019} or can complement the content~\cite{de2018}.

However, people with visual impairments have been largely underserved in receiving the benefit of ``using vision to think'' that visualizations offer. Not being able to understand visualizations may result in aggravating information inequality of people with visual impairments. A few assistive technologies have been proposed to bridge this gap over the past years, attempting to make visualizations accessible to a broader audience. For example, popular approaches include tactile visualizations (e.g.,~\cite{yang2020tactile,guinness2019robographics}) that use embossed surfaces to present visualizations and data sonification (e.g.,~\cite{sakhardande2019comparing,zhao_data_2008}) that maps the data to various dimensions (e.g., pitch, volume) of sounds. However, integrating these technologies into the online environment, in which we frequently consume data, is a major challenge.

Communicating visualizations with text can be a practical way to convey the information in visualizations, especially in web environments. For example, HTML supports many ways to add invisible text fields near graphics to allow people with visual impairments to access the description of graphics through screen readers. These descriptions used to \textit{replace} graphics are known as alternative text (alt text). Alt text serves as a textual substitution of a graphic that describes the visual components within it. This helps people to construct the graphic mentally when they have no access to the graphic. Several organizations provide guidelines to create alt text for visualizations. However, the guidelines do not offer rationales and empirical evidence for their recommendations. Thus, it remains unclear whether the suggested ways of generating alt text align with what people with visual impairments need or how effective the recommendations are.

In this work, we investigate how to formulate alt text for visualizations to best serve people with visual impairments. As a first step, we surveyed current guidelines on generating visualization alt text. Next, we reviewed current practices to understand to what extent people follow the guidelines and what other approaches people use when formulating alt texts. As we could not find many examples of alt texts generated for visualizations from major news outlets, we collected alt texts from academic papers. We focused on alt texts written by people who have expertise in visualizations and graphics (IEEE VIS \& TVCG), have knowledge of accessibility (ACM ASSETS), or are often prompted to write alt texts for visualizations (ACM CHI as the submission system encourages authors to submit alt texts for figures). We found that people use diverse strategies, ranging from providing brief explanations of the charts to listing all data points. The observed practices are only partially aligned with the surveyed guidelines.

To evaluate the surveyed guidelines and further probe information needs for visualization alt texts, we interviewed 22 participants with visual impairments. In the study, participants examined visualizations by reading the alt text formulated with several strategies through their screen readers. Then, participants were asked to share their preferences and opinions on the strategies and provide insights toward the ideal visualization alt texts. Our findings support many of the surveyed guidelines, such as the need to provide information about the chart type, axes, access to data points, and description of data trends. However, contrary to guidelines, some participants want to enrich their mental picture of the visualization by knowing visual attributes used in visualization, such as color encoding. This is in line with another finding of the study -- that participants wish to ``visualize'' visualizations in their head while listening to alt texts. This finding implies that alt texts should describe the necessary visual components to fill the gap in their imagination and ease the construction of an image of the visualization. Additionally, our study findings provide evidence to guide how an author should generate alt texts for their visualizations, including which component to mention first, how an author should describe the trend in the visualization, and what language tone the author should maintain. 

Our contribution is three-fold:
\begin{enumerate} [topsep=-0mm,itemsep=0mm,partopsep=0ex,parsep=0.1ex]
    \item We survey and summarize existing guidelines for the generation of visualization alt text. We also analyze current practices to demonstrate how aligned they are with the guidelines and identify other commonly used strategies.
    \item We report findings from an interview study with 22 people with visual impairments that provide insights into the properties of good alt texts and into the role of visualizations for people with visual impairments to support people with visual impairments in the context of reading online news.
    \item We propose actionable recommendations for generating informative alt texts along with rationales and suggest system-level improvements (e.g., structures of HTML, multimodal integration) informed by the study to better support visualization interpretation for people with visual impairments.
\end{enumerate}

\section{Background}
People with visual impairments are a large population that includes people who are blind and have low vision. According to the World Health Organization, about 285 million people in the world have visual impairments, of whom 39 million are entirely blind~\cite{WHO}. Due to their vision loss, this population faces difficulties in understanding graphics, which hinders their education and employment opportunities.    

With the advances of the Internet and computer technology, numerous physical readings and learning materials have been digitized into web pages or other digital formats (e.g., pdf, doc). To enable people with visual impairments to access digital information, screen readers have been developed. Screen readers are software programs that scan the text on a screen and communicate the information to people with visual impairments via speech (some also have braille output via a braille display)~\cite{AFB}. Users control which components in the page the screen reader communicates through several key combinations on the keyboard or a touch gesture on a smartphone.
Most mainstream computer or smartphone systems have embedded screen readers. Some examples include VoiceOver for Mac and iOS \cite{VoiceOver}, Narrator for Windows 10 \cite{Narrator}, and TalkBack for Android \cite{TalkBack}. There are also standalone screen reader softwares, such as JAWS~\cite{JAWS} and NVDA~\cite{NVDA}.    

While making text more accessible, screen readers cannot interpret images for people with visual impairments. Alternative text is the major source that screen readers rely on to communicate graphics. Alternative text, also known as ``Alt text,'' was first introduced to the HTML 2.0 specification in 1995~\cite{HTML2}. Alt text is an invisible text block that can be inserted as an \texttt{alt} attribute of the \texttt{<img>} tag. The original purpose was to inform web viewers of an image's content when the image could not show properly. Nowadays, alt text is commonly used by people with visual impairments to access image content via screen readers. Adding alt text to images has thus become one of the most important principles for web accessibility~\cite{AltText}. Similarly, the \texttt{longdesc} attribute in an \texttt{<img>} tag is used to provide a more lengthy explanation about non-text elements~\cite{longdesc}. URLs, a location in the page referred by id or class, or text can be part of \texttt{longdesc}. Even though \texttt{longdesc} is deprecated in HTML5,  WCAG still recommends it to be provided for complex graphics.

For other graphic formats such as SVG elements, the \texttt{<title>} and \texttt{<desc>} tags can also be used to convey the contents of visualizations in an easily accessible textual form. However, compatibility issues with some screen readers have been reported~\cite{bugAria,bugjaws}. Beyond web pages, alt text features have been extended to various digital document formats (e.g., docx, pptx, pdf). Commercial content-editing and viewing software, such as Microsoft Word and Adobe Acrobat, also enables users to add alt text to images, thus making digital documents more accessible to people with visual impairments~\cite{MS-AltText,Adobe-AltText}.

\subsection{Visualization accessibility for visually impaired people}
Several strategies have been explored for communicating visualizations to people with visual impairments by relying on senses other than vision. One popular approach is sonification~\cite{choi2010digitizer,Walker2010,sakhardande2019comparing}, which seeks to encode the data of a chart into various dimensions (e.g., pitch, volume, pan) of an audio signal. Several studies have shown that audio charts can communicate patterns in the data effectively~\cite{choi2010digitizer,sakhardande2019comparing}. Data sonification has been used on various visualizations such as line graphs, bar graphs, and maps ~\cite{zhao_data_2008, demir_2010, heuten_interactive_2006}. Tactile visualizations realized via haptic feedback, braille display, or embossed prints are another popular approach to enhance accessibility for people with visual impairments~\cite{watanabe2018effectiveness, Engel2017,engel2018user,engel2017analysis,engel2017improve,paneels2009review,yang2020tactile,Fusco2015,hu2015,Engel2019}. Yet, the low-resolution nature of tactile visualizations limits their use for more complex visualizations~\cite{engel2017improve}. Another recent exploration suggests that olfaction can be use to perceive data~\cite{patnaik2018}. To overcome the limitation of one modality, multiple sensory inputs can also be combined to complement each other~\cite{hahn2019,goncu2011,taibbi2014,Lazar2013,Wall2006a,gardner2006scientific,landau2001development}.

Summarizing visualizations in textual form is another way to convey visual information to people with visual impairments~\cite{elzer2007,Moraes2014,granz2013}. EvoGraphs is a jQuery plugin that enables creating visualizations that can be read by screen readers~\cite{Sharif2018}. EvoGraphs automatically formulates an alt text that contains each data point and some representative values of the data (e.g., max, min, mean). However, these systems often focus on one or two basic chart types, such as bar and line charts. While not aiming at supporting people with impairment specifically, several techniques have been proposed to automatically generate captions for visualizations~\cite{fasciano1996,corio,mittal,kafle2018,kahou2017,chen2019,obeid2020,wu2010,burns2013}. A more thorough survey on accessible visualizations can be found~\cite{kim2021acc}.

\subsection{Study on alternative text for images}
Unlike visualization alt texts, methods to generate alt texts for images (i.e., image description\footnote{Conventionally, the community concerning alternative texts for images refers the alt texts more generally as image descriptions or image captions}) have been explored extensively. As presented in~\cite{Stangl2020}, alt text generation can be categorized in three groups: human-generated (e.g.,~\cite{bigham_vizwiz_2010}), computer-generated (e.g.,~\cite{vinyals_show_2015, wu_automatic_2017}), and hybrid approaches (e.g.,~\cite{morris2018, salisbury_toward_2017, salisbury_evaluating_2018}). While many efforts have been put into investigating formulating and evaluating image descriptions, a large proportion of images found online lack alt text descriptions~\cite{gleason_its_2019}. 

Recommendations for alt text formulation depend on the purpose of the image, presented objects, individuals, active motions, location, colors, and emotions~\cite{petrie2005}. Studies in image alt text often focus on social network contexts where people share their photos actively~\cite{bennett2018, morris2016, voykinska2016, zhao2017}. Updated guidelines for social networking sites suggest containing elements such as the number of people and facial expressions~\cite{wuask2016}. Recent research demonstrates that people with visual impairments have different needs based on the context where they encounter the images (e.g., social network, e-commerce)~\cite{Stangl2020}. 
\section{Phase 1: Guidelines for Visualization alt texts}
\label{sec:phase1}
To understand the current standard, we collected the existing guidelines for generating visualization alt texts. Since the goal of alt texts is to replace the graphics, most guidelines address how to describe visual components in visualizations to support people to mentally construct visualizations.

\subsection{Guideline collection}
We collected guidelines via Google search using relevant keywords such as visualization accessibility, visualization alternative text, etc., resulting in 31 postings. We excluded accessibility guidelines that were not explicitly related to blind or low vision individuals (e.g., guidelines for color blind people), guidelines not focused on formulating alt texts, or guidelines created by citing other guidelines. This process resulted in four sets of guidelines, namely the WCAG guidelines~\cite{wcagcompleximages}, the Penn State's accessibility guidelines~\cite{penn_state_2018}, the Diagram Center's guidelines~\cite{diagram_center_2020}, and CFPB's guidelines~\cite{cfpb}. The Web Content Accessibility Guidelines (WCAG) 2.1 is a widely used collection of recommendations for increasing the accessibility of Web content~\cite{w3c2018}. WCAG provided guidelines for visualization alt text through the W3C Web Accessibility Initiative~\cite{wcagcompleximages}. Accessibility at Penn State complemented the WCAG guidelines with specific examples of a few different chart types~\cite{penn_state_2018}. The Diagram Center is a research development center that provides specific guidelines for various types of graphics such as charts~\cite{diagram_center_2020}. 
Finally, CFPB Design System is an open-source resource for teams at the Consumer Financial Protection Bureau (CFPB) that helps teams produce accessible products~\cite{cfpb}.

\subsection{Findings}
\begin{table}[t!]
\resizebox{\columnwidth}{!}{
\begin{tabular}{p{0.18\linewidth} | p{0.82\linewidth}}
\hline
\textbf{Source}             & \textbf{Summary}
\\ \hline
WCAG~\cite{wcagcompleximages} & Complex images, which includes graphs, charts, and maps, must include a two-part text alternative. The first part is a short description that identifies the complex image. The second part is a long description that is a textual representation of the essential information. The WCAG 2.0 Guideline 1.1.1 outlines that all non-text contents must have a text alternative to present the equivalent purpose.
\\ \hline
Accessibility at Penn State~\cite{penn_state_2018} & If the data in a complex image is essential, a text description of the image must be provided. In appropriate cases, a numeric table representing the chart data will provide additional accessibility.
\\ \hline
Diagram Center~\cite{diagram_center_2020}              & Charts and graphs must be converted into accessible tables; a brief description and a summary, if needed, should be provided.  Additional information such as the title and axis labels should be included as well. However, visual attributes such as color are not necessary to include unless there is an explicit need for them. 
\\ \hline
CFPB~\cite{cfpb} & The alt texts should include one sentence of what the chart is and the chart type. There should also be a link to a CSV or another machine-readable data format with the raw data. Moreover, the data must have descriptive column labels. Take into consideration that screen readers do not let users skip or speed up while reading alt texts.
\\ \hline
\end{tabular}
}
\caption{Brief summary of each guideline source.\label{table:guidelines}}
\end{table}

Table~\ref{table:guidelines} summarizes the four guidelines. These guidelines offer recommendations for generating alt texts to visualizations, including the structure of alt text and which components to include.

\textbf{Structure of alt text.}
WCAG recommends two-part alt texts for complex visualizations, with the first part containing a short description of the chart and the second part a more detailed description~\cite{wcagcompleximages}.
WCAG suggests not to include the long description in an alt text. Instead, they propose various ways to incorporate long descriptions, for example, by specifying a link to the long description with an \texttt{<a>} tag adjacent to the chart or using the \texttt{longdesc} attribute. According to WCAG, the short description should also indicate how/where to access the long description.

Concerning length, different guidelines prescribe varying principles. While WCAG~\cite{wcagcompleximages} encourages longer descriptions as visualizations contain substantial amount of information, other guidelines~\cite{diagram_center_2020,penn_state_2018,cfpb} suggest one or two sentences. For example, CFPB asserts that alt texts must be short yet descriptive since screen readers may not allow users to skip or speed up while reading alt text~\cite{cfpb}.


\begin{table*}[h!]
 \resizebox{\linewidth}{!}{
\begin{tabular}{|l|l|l|l|l|l|l|l|l|l|l|}
\hline
\textbf{Pid} & \textbf{Age} & \textbf{G.} & \textbf{Edu.} & \textbf{Occupation}    & \textbf{Diagnosis}                               & \textbf{Onset Age (year)} & \textbf{Light Perception} & \textbf{Screen Readers} & \textbf{Years Used}  \\ \hline
P1           & 22           & M           & H.S.          & Customer service       & Retinal detachment                               & 0              & N                         & NVDA                    & 5           \\ \hline
P2           & 38           & F           & M.S.          & Accessibility tester   & Leber hereditary optic neuropathy                & 0              & N                         & JAWS                    & 20          \\ \hline
P3           & 24           & M           & B.S.          & Student                & Juvenile macular degeneration                    & 6              & Y                         & VoiceOver                    & 10          \\ \hline
P4           & 22           & F           & H.S.          & Student                & Astrocytoma resulting in optic nerve compression & 9              & N                         & JAWS                    & 11          \\ \hline
P5           & 46           & F           & M.A.          & Unemployed             & Albinism, Glaucoma                               & 0              & Y                         & JAWS                    & 21          \\ \hline
P6           & 31           & F           & B.A.          & Writer                 & Retinopathy of prematurity                       & 0              & N                         & JAWS                    & 20          \\ \hline
P7           & 26           & M           & B.A.          & Unemployed             & Retinopathy of prematurity                       & 5              & N                         & JAWS                    & 14          \\ \hline
P8           & 35           & F           & B.A.          & Customer service       & Leber hereditary optic neuropathy                & 0              & N                         & JAWS                    & 20          \\ \hline
P9           & 23           & F           & H.S.          & Student                & Retinopathy of prematurity                       & 0              & N                         & JAWS                    & 12          \\ \hline
P10          & 26           & F           & M.A.          & Policy analyst         & Retinopathy of prematurity                       & 0              & N                         & JAWS                    & 6           \\ \hline
P11          & 20           & M           & A.A.          & Developer              & Leber hereditary optic neuropathy                & 0              & Y                         & VoiceOver                    & 10          \\ \hline
P12          & 31           & M           & B.S.          & Usher                  & Glaucoma                                         & 0              & Y                         & JAWS                    & 10          \\ \hline
P13          & 32           & F           & H.S.          & Information specialist & Glaucoma                                         & 0              & Y                         & JAWS, NVDA              & 7           \\ \hline
P14          & 25           & M           & H.S.          & Student                & Glaucoma and Peters Anomaly                      & 4              & N                         & JAWS, NVDA              & 7           \\ \hline
P15          & 25           & F           & B.S.          & Student                & Microphthalmia                                   & 0              & N                         & VoiceOver                    & 16          \\ \hline
P16          & 26           & F           & J.D.          &  Claims specialist & Leber's congenital amaurosis                         & 0              & Y                         & JAWS, NVDA              & 18          \\ \hline
P17          & 24           & F           & M.A.          & Speaking employment    & Leber's congenital amaurosis                     & 0              & Y                         & JAWS                    & 12          \\ \hline
P18          & 31           & F           & B.A.          & Unemployed             & Optic Atrophy                                    & 0              & Y                         & JAWS                    & 21          \\ \hline
P19          & 27           & M           & B.S.          & Customer service       & Optic Atrophy                                    & 8              & N                         & VoiceOver                  & 17          \\ \hline
P20          & 30           & F           & H.S.          & Student                & Retinopathy of prematurity                       & 0              & Y                         & JAWS                    & 25          \\ \hline
P21          & 29           & F           & M.A.          & Teacher                & Genetic mutation                                 & 0              & Y                         & JAWS                    & 5          \\ \hline
P22          & 37           & F           & M.S.          & Technology specialist  & Retinopathy of prematurity                       & 0              & N                         & JAWS                    & 25          \\ \hline

\end{tabular}
}

\caption{Demographics of participants. \textit{Pid=Participant ID}. \textit{G=Gender} (M=Male, F=Female). \textit{Edu=Education} (H.S.=High School, B.S.=Bachelors of Science, B.A.=Bachelors of Arts, M.A.=Masters of Arts, M.S.=Masters of Science, J.D.=Doctor of Jurisprudence, A.A.=Associates).}
\vspace{-2mm}
\label{table:participants}
\end{table*}

\textbf{Components in alt text.} 
Alt text should provide a meaningful and informative description of a visualization that is sufficient for a reader to understand its content~\cite{cfpb}.
To support this goal, CFPB and the Diagram Center's guidelines encourage alt texts to include a one-sentence summary of the chart, chart type, and axis labels~\cite{cfpb,diagram_center_2020}.
WCAG also suggests mentioning all visually presented scales and values~\cite{wcagcompleximages}. According to the Diagram Center's guidelines, visual attributes (e.g., the color of a bar or line types) don't need to be explained unless there is a special need~\cite{diagram_center_2020}. Some guidelines also encourage to include a summary of the data trends~\cite{diagram_center_2020,cfpb,wcagcompleximages}.

\textbf{Data tables.} 
For complex visualizations, the conventional alt text provided via the \texttt{alt} attribute in HTML may not suffice to provide the information needed. In addition to a text summary, a data table can further enhance accessibility to a visualization. Both CFPB's and Diagram Center's guidelines suggest including a link to the data represented in the visualization in a format accessible to screen readers,  e.g., CSV or other machine-readable formats~\cite{cfpb, diagram_center_2020}. Also, tables should contain descriptive column labels~\cite{cfpb}.

\subsection{Summary}
Overall, the guidelines emphasize the need for descriptive and succinct language in alt texts, including information on the chart type, axes, and data trends. Some visual attributes (e.g., color) may be omitted unless there is an explicit reason. Many sources echoed that formatted tables are essential in understanding visualizations, even though they are not a part of alt texts. While these guidelines provide a useful starting point, they lack rationales for why each component should or should not be included. Furthermore, the guidelines do not reference empirical evidence of how they support the needs of people with visual impairments.

\section{Phase 2: Analyzing Current Practices}
In addition to surveying guidelines, we inspected current practices to further understand the status quo. The goal is to examine whether authors follow the guidelines, and if not, to identify which strategies authors use when formulating alt text for visualizations. 

We sought to collect alt text and visualization pairs from online media outlets as this is a common way in which the general public consumes visualizations. We sampled 30 visualizations from three major news outlets (NYT, The Washington Post, FiveThirtyEight) who frequently published visualizations alongside articles. However, we found no alt texts associated with the sampled visualizations (no \texttt{alt} attribute if the visualization was presented as an image element, nor \texttt{desc} tag if the visualization was presented as an SVG element). Hence, we shifted our approach to obtain examples from academic publications. Specifically, we collected alt text and visualization pairs from IEEE VIS \& TVCG, ACM ASSETS, and ACM CHI over the last two years (i.e., 2019, 2020). We chose these three publication venues to favor authors with expertise in visualizations and graphics (IEEE VIS \& TVCG), with knowledge of accessibility (ACM ASSETS), or authors that are often prompted to write alt texts for visualizations (ACM CHI as the submission system encourages authors to submit alt texts for figures). 

\subsection{Data collection \& data preliminary}
We downloaded the academic publications in a PDF format from the IEEE Xplore digital library and ACM digital library. We then converted the PDFs to accessible text containing the embedded alt texts using Adobe Acrobat. Next, we used a custom Python script to identify and extract alt texts from the accessible text. We also downloaded the figures from Semantic scholar and mapped them to the extracted alt texts by matching both the publication DOI and the figure number.

We collected total of 2,278 publications (VIS\&TVCG:723, ASSETS:95, CHI:1,460) with 7,493 figures (VIS\&TVCG:2,518, ASSETS:281, CHI:4,694). Among those figures, 40\% contained alternative text descriptions (VIS\&TVCG:0\%, ASSETS:65\%, CHI:51\%). To filter only visualizations (e.g., bar, line, area chart, scatterplot, boxplot etc.) from plain images, one researcher reviewed the figures manually. After filtering, we removed the alt texts that only contain the figure number (e.g., ``Figure 1'') or random placeholders (e.g., ``abc''), resulting in 0 alt text-visualization pairs from the IEEE VIS \& TVCG collection, 89 pairs from the ASSETS collection and 752 pairs from the CHI collection.

For each pair in the aggregated collection, we marked whether the alt texts mentioned summary, chart type, axes, data trends, visual attributes (color and shape), and data points. 
Two researchers annotated all alt text-visualization pairs independently and discussed them together to resolve disagreements (there were around 5\% of disagreements initially). We also calculated the length of each alt text using NLTK's sentence tokenizer~\cite{nltk}.  

\begin{figure}[h!]
    \centering
    \includegraphics[width=\linewidth]{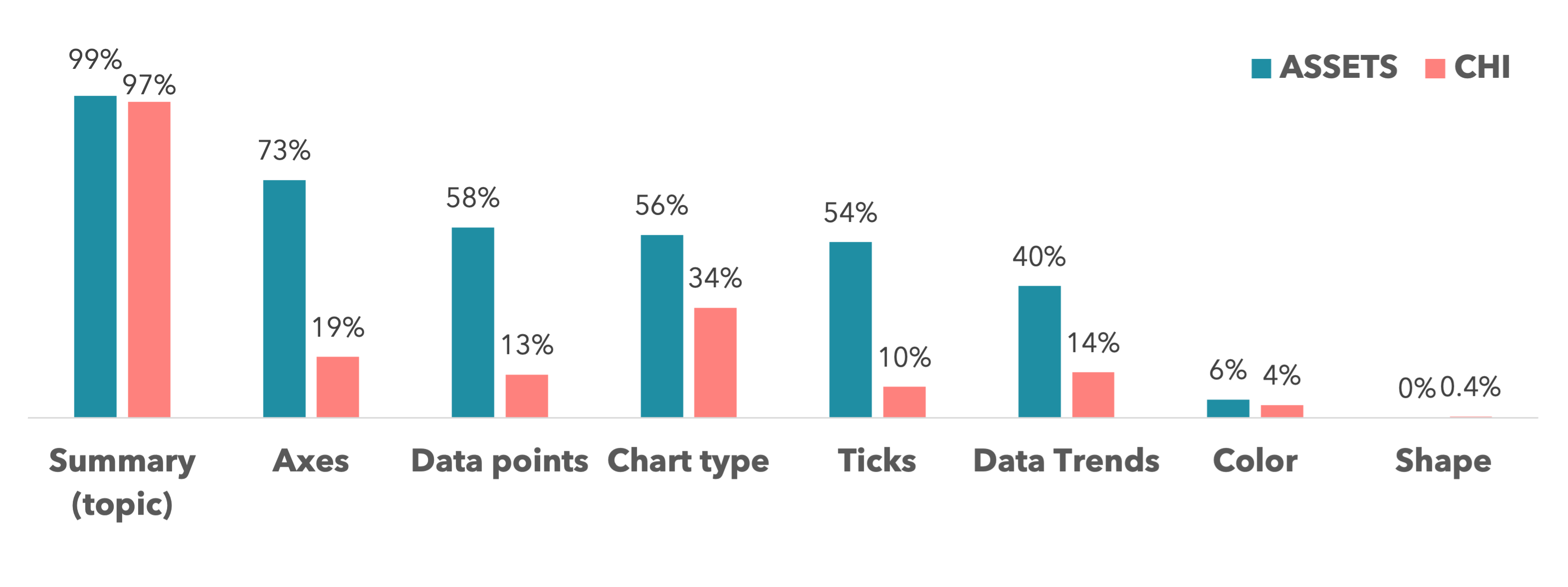}
    \vspace{-8mm}
    \caption{The frequency of each component that appears in the alt texts from the ASSETS and CHI collections.}
    \label{fig:phase2_stat}
\end{figure}

\subsection{Results}
Figure~\ref{fig:phase2_stat} shows the frequency that each component appears in alt texts for the ASSETS and CHI collections. Almost all alt texts include a summary that describes the visualization's overall topic (e.g., ``Input speed of three techniques''). The second most common component for the ASSETS collection was axes information (65 out of 89 alt texts, 73\%). In the CHI collection, chart type was mentioned 34\% of the time (254 out of 752 alt texts). In both collections, visual attributes (e.g., color and shape) were the least common among all components.

\begin{figure}[h!]
    \centering
    \hspace{-10mm}
    \includegraphics[width=\linewidth]{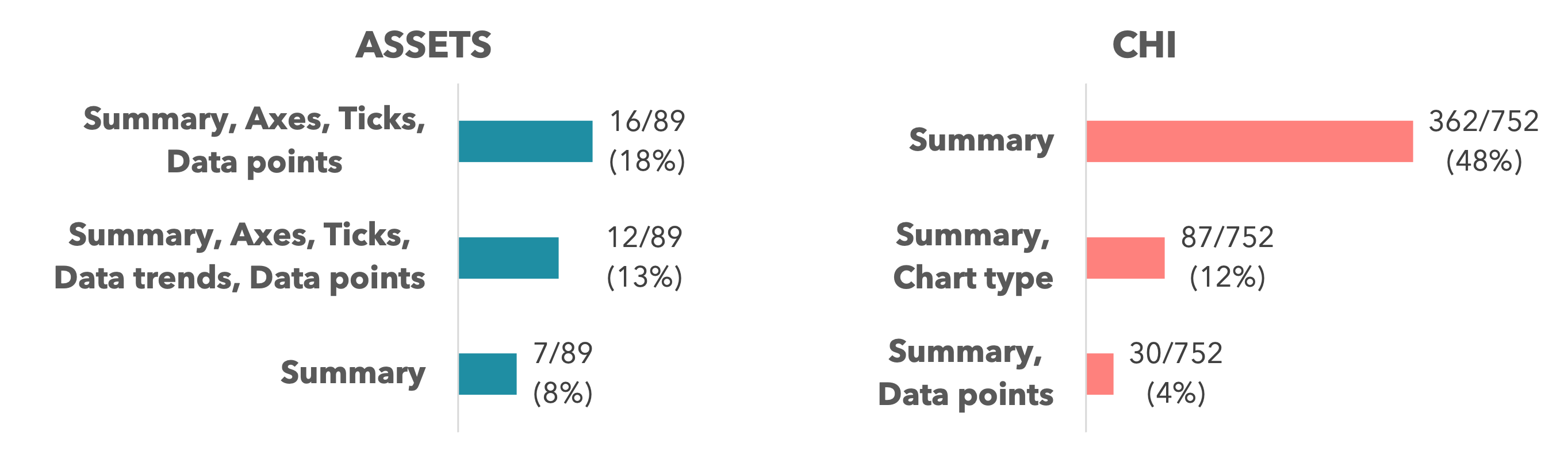}
    \vspace{-5mm}
    \caption{The frequency of the top-3 most common combinations of components used in alt texts from the ASSETS and CHI collections.}
    \label{fig:phase2_top3}
\end{figure}

We also analyzed which combination of components was most used when formulating alt texts (Fig.~\ref{fig:phase2_top3}). A summary of the visualization together with axes, ticks, and data points was the most common combination (18\%) in the ASSETS collection. The second most common combination also included data trends in addition to the summary, axes, ticks, and data points (13\%). Finally, alt texts that only provide a summary were the third most common (8\%). In the CHI collection, a summary alone was the most common alt text type (48\%), followed by a summary with the chart type (12\%) and a summary with data points (4\%). 

The average length was 3.5 sentences (Median=3, SD=3.3). Alt texts from ASSETS (M=4.4, SD=2.9) were slightly longer (t=-2.8, p<0.01) than those in the CHI collection (M=3.5, SD=3.5).

\subsection{Summary}
The analysis shows that the authors do not entirely follow current guidelines. For example, the authors only mention the actual data points, chart type, and data trends about half of the time in the ASSETS collection and less than a third of the time in the CHI collection. While alt texts from the ASSETS collection are better aligned with the guidelines, only half of the alt texts contain the chart type (e.g., ``this chart describes'' instead of ``this bar chart describes''). Without mentioning the chart type, visually impaired individuals may struggle to imagine the depicted visualization. The CHI collection seriously lacks visualization accessibility since 48\% of the time, the only available information is the summary. Length-wise, the collected alt-texts were longer on average than the guidelines often prescribe, but the numbers were highly varied. 

All 30 visualizations we sampled from major news outlets did not contain alt texts associated with the visualizations. Since visualizations are a critical part of information consumption, this observation is alarming. Similarly, the IEEE VIS \& TVCG collection did not include any text alternatives to the figures. To exclude the possibility that the script couldn't detect alt texts from the collection, we randomly chose 100 papers and manually checked them through Adobe Acrobat. 
We urge the IEEE VIS community to employ a system that encourages the use of alt texts to make science more accessible to people with visual impairments.

\section{Phase 3: Semi-Structured Interview}
To evaluate the guidelines and further understand the needs of people with visual impairments concerning visualization alt texts, we conducted semi-structured interviews. Specifically, the goal of the study is to identify empirical evidence to motivate guidelines and further derive implementation in alt text formulation for visualization. We examined participants' behaviors in the news consumption scenario, which we believe is one of the frequent encounters of visualizations online for the general public. 

\subsection{Method}
\subsubsection{Participants}
We solicited study participation by circulating an IRB-approved flyer on listservs hosted by organizations serving the blind and low vision population. Our recruitment criteria were designed to recruit participants who are 1) at least 18-year-old, 2) legally blind, and 3) using screen readers daily. We received 150 responses from potential participants, contacted participants for the interview on a first-come-first-served basis, and recruited participants until we saturated the findings~\cite{alroobaea2014}.
The final pool of participants consisted of 22 participants (15 female, 7 male), whose ages ranged from 20 to 46 (M=28.6, SD=6.2, Table~\ref{table:participants}). Among 22 participants, 21 were blind, and 1 had low vision. All interviews were conducted via Zoom. Each interview lasted on average 60 minutes with a standard deviation of approximately 10 minutes. We compensated their participation with a \$20 Visa gift card.  

\begin{figure*}[h!]
    \centering
    \includegraphics[width=\linewidth]{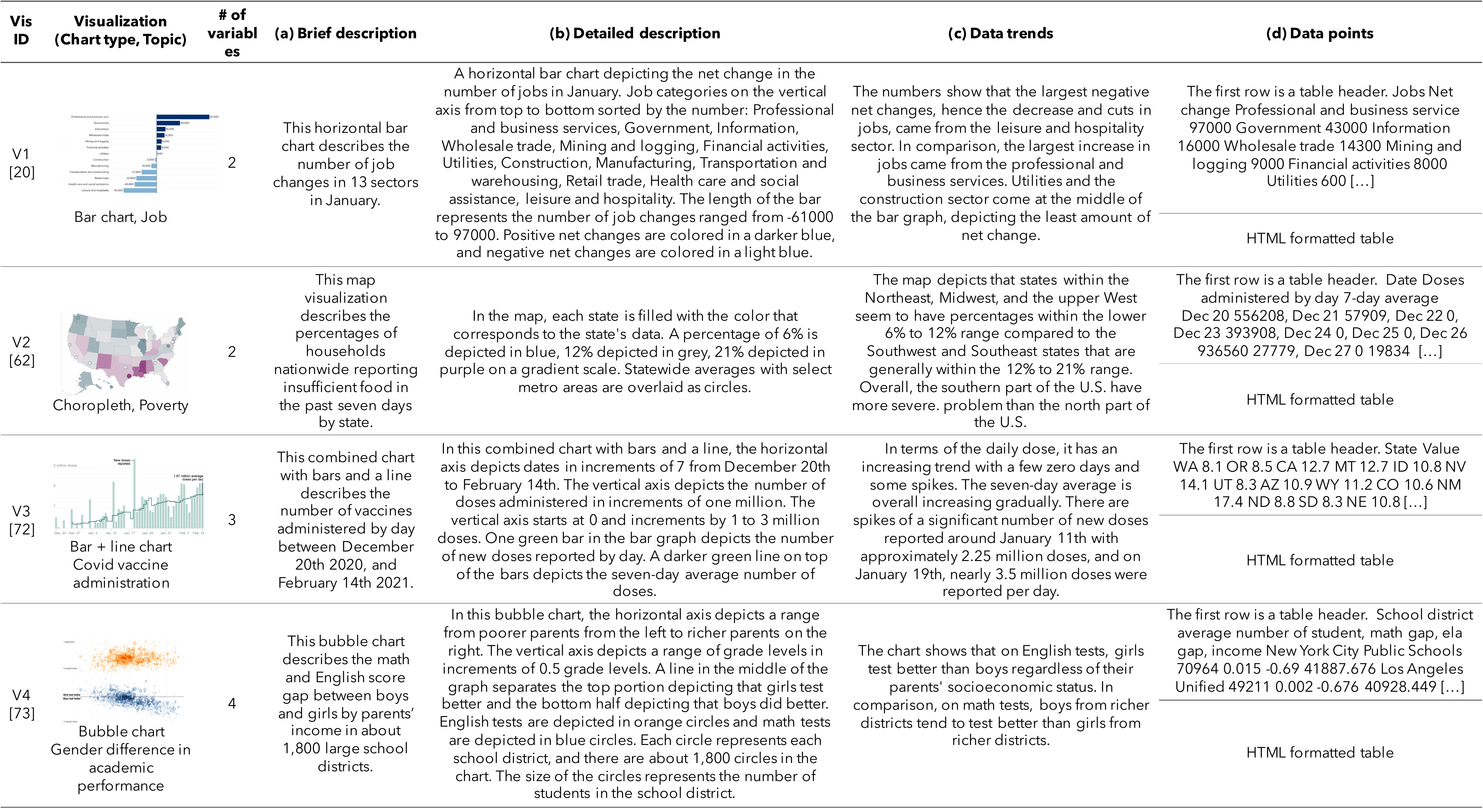}
    \caption{Study stimuli. Each participant saw all of the visualizations parked with a different style of alt text.   \phantom{V1~\cite{news1}, V2~\cite{news2}, V3~\cite{news3}, V4~\cite{news4}}}
    \label{fig:stimuli}
    \vspace{-5mm}
\end{figure*}

\subsubsection{Study stimuli}
To provide the context of a real-world visualization reading scenario, we prepared four articles from online news covering different topics. We included multiple chart types to maximize participants' exposure to different components (e.g., bar, line) of the visualization, as well as the number of variables encoded to enable participants to examine additional encodings (e.g., color), to expose them to varying complexities. Figure~\ref{fig:stimuli} shows the selected visualizations. 

In creating the study stimuli, we shortened the original article text to only one or two paragraphs related to the visualization, so as to conduct the study in a reasonable amount of time (e.g., 1 hour). 

Three out of the four visualizations (V2, V3, V4) were in an SVG format. For SVG elements, \texttt{desc} tags serve to place alt texts, equivalent to the \texttt{alt} attribute for an image. However, due to potential comparability issues of \texttt{desc} tags with some screen readers~\cite{bugjaws}, we converted the three visualizations to a bitmap image and added an alt text via the \texttt{alt} attribute. 

For each of the four visualizations, we formulated alt texts in four different styles informed by Phase 1. The four styles contain different visualization components and different specificity mentioned in the guidelines to prompt participants to think about their preference when some components and the specificity are present and absent (Fig~\ref{fig:stimuli}):

\begin{itemize}[leftmargin=3mm]
    \setlength\itemsep{-0.3mm}
    \item  Brief description (Fig.~\ref{fig:stimuli}a): Since several guidelines from Sec.~\ref{sec:phase1} emphasize to have a short description~\cite{wcagcompleximages,cfpb}, we formulated a brief description. We first described the type of visualization~\cite{cfpb} followed by a summary of the data that the visualization represents.
    \item Detailed description  (Fig.~\ref{fig:stimuli}b): A few guidelines encourage to include more detailed aspects of the visualization~\cite{wcagcompleximages,diagram_center_2020}. In this style, we first described the type of chart followed by a summary, a depiction of visualization elements including axes, range, marks, and mapping. To evaluate the guideline discouraging description of visual attributes (e.g., color, shape)~\cite{diagram_center_2020}, we included them in this style of description, when applicable. 
    \item Data trends  (Fig.~\ref{fig:stimuli}c): Data trends are another component that many guidelines suggested to include in alt texts~\cite{diagram_center_2020,cfpb,wcagcompleximages}. We described the overall trends that were visually apparent.
    \item Data points (Fig.~\ref{fig:stimuli}d): All the guidelines we surveyed highlight the importance of having access to raw data~\cite{penn_state_2018,cfpb, diagram_center_2020,wcagcompleximages}. To validate the emphasis on having a separate table element in the guidelines, we created two different ways to access the data. The first version includes data points within the alt texts, which sequentially read out the raw data. The second version includes the data in a formatted HTML table. This allows the participants to navigate the table with their screen readers. In this case, we indicated that a table could be found below within the alt text.
\end{itemize}

We formulated four different alt texts styles for each of the four chosen visualizations, resulting in 16 different stimuli. All participants saw all of the four articles paired up with one of each of the alt text styles. We counterbalanced the pairing between the articles and styles, and randomized the order of presenting the articles. Specifically for the Data points style, half of the participants saw the data points in the alt text and the other half saw the data points in an HTML format. 

\begin{figure}[h!]
    \centering
    \includegraphics[width=\linewidth]{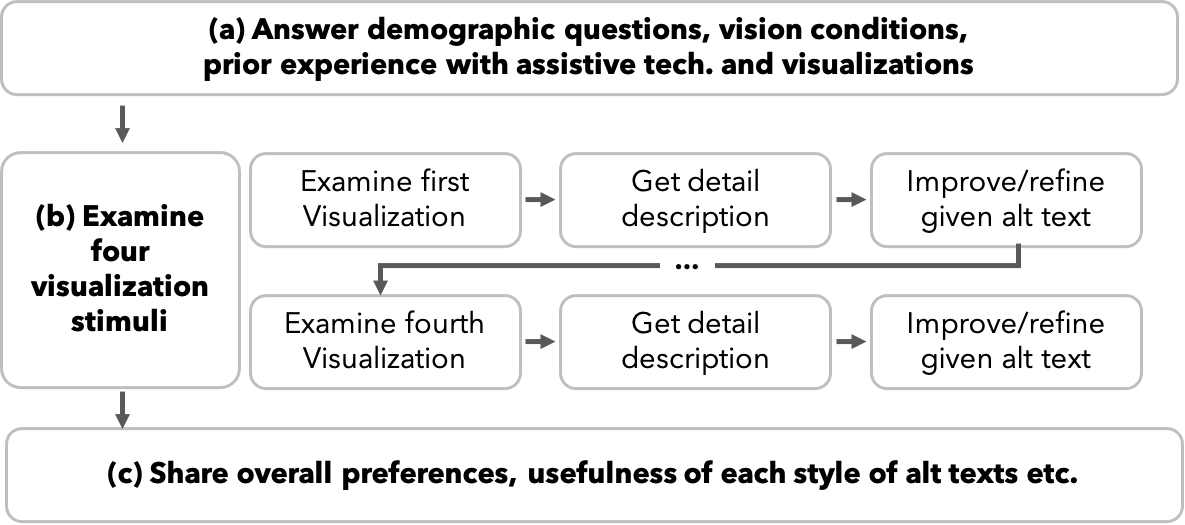}
    \caption{Study procedure. We asked participants their demographic information, conditions, and prior experience. After that, participants examined each visualization once at a time by reading the alt text with their screen readers. After each visualization, we gave them a full description of the visualization and asked them to improve the alt text based on their information needs. After examining all of the visualizations, we asked questions regarding their preferences and needs toward alt texts.}
    \label{fig:procedure}
\end{figure}

\subsubsection{Procedure}
Figure~\ref{fig:procedure} shows the overall procedure of the interview. We first asked participants demographic questions, including their age, gender, education level, occupation, as well as their vision conditions, including their diagnosis, onset age, and remaining vision. Then, we asked about their experience with assistive technologies. Specifically, we asked what assistive technologies they have been using and how long they have been using them. We also asked participants how often they encountered visualizations and the situations they encountered them. 

Next, participants were prompted to open the URLs presenting the study stimuli that were sent just before the study started through email. At that point, we asked them to share their screen to observe which part of the study stimuli they were interacting with. While the participants interacted with the stimuli, we asked them to express what they were doing every step of the way (think aloud). 

We then asked participants what they learned about the visualization and how they learned that information. The example questions include: What do you think that the chart is about? How does the alt-text help you learn about the chart? Besides alt-text, have you used any other strategies to understand the chart? After each visualization, we gave a full description of the visualization, consisting of the detailed description, data trends, and the summary of data points (Fig.~\ref{fig:stimuli}b, c, d). We then asked the participants how they would regenerate or refine the alt texts and which aspect of the given alt texts was helpful to envision the visualization and the underlying data.

After examining all four visualizations, we asked their preference regarding length, contents, and usefulness of alt texts. 
Participants were also asked about the role of visualizations in understanding the article, how they compare the alt texts requirements between images and visualizations, and the perfect alt text system that they envision. 

\subsubsection{Analysis}

We first transcribed the recorded study sessions. We conducted a thematic analysis, a method to identify themes within the data~\cite{braun2006}.
Three researchers coded three participants' transcripts independently and discussed them together to create a consolidated codebook. Then, we categorized the codes in the codebook based on the emerging themes using affinity diagrams and axial coding. The discussion resulted in 12 themes (e.g., Prior experience, Visual components) and 152 codes. One researcher coded the rest of the transcripts using the codebook.

\subsection{Findings}
We present the findings from our thematic analysis of the semi-structured interview conducted on 22 participants. 

\subsubsection{Prior Experience with Visualizations}

\textbf{Context.}
Thirteen participants stated they read visualizations in academic settings. P20 stated, \textit{``I encounter visualizations a few times a week at school.''} The participants that read visualizations in an academic setting expressed that visualizations raise questions such as, \textit{``What is this showing? What do these numbers represent?''} and lack \textit{``any descriptive text except for possibly a caption,''} as stated by P3.

Participants also found visualizations while seeking information on the web. As stated by P17, \textit{``There's a lot of charts and graphs when looking for information about current events. The charts and graphs show what areas have increased cases and other information.''} In addition, we observed that another common situation where visualizations appear is in professional settings. As noted by P6, \textit{``I encounter charts a lot when I'm observing survey data,''} and also by P8, \textit{``When I am at work there's some charts [that are related to my work].''}

\textbf{Frequency.} In terms of frequency, two participants mentioned that they encounter visualizations on a daily basis, nine participants mentioned more than three times a week, and eleven participants mentioned once a week.

\textbf{Assistive technology.} Braille displays and tactile materials are a common medium used by participants to read visualizations, especially in academic settings. 15 out of 22 participants explicitly mentioned using them. For example, P10 stated she has been reading visualizations \textit{``through tactile braille, embossed braille, and hardcopy materials.''} P11 shared, P11 \textit{``for images specifically I use a screen reader. But the main way that I do this [visualization] is through tactile.''} Many participants also expressed their familiarity and extended use of braille displays. For example, P15 mentioned that \textit{``I read braille, and I've been using that my entire life,''} and P10 noted that \textit{``I have been reading braille since I was in kindergarten, so I've been reading braille for a long time.''}

We observed that familiarity with tactile visualizations allowed participants to better understand particular visualizations, such as bar charts.
As stated by P10, \textit{``We've grown up learning those concepts and know like what a bar chart is, I know what a histogram is.''} 
However, some uncommon types of charts that they have never experienced with a braille display might lead to confusion, as noted by P15: `\textit{`I’ve never seen a bubble chart before.''} 

\subsubsection{The Motivation to Understand Visualizations}

The motivation to understand visualizations spanned several reasons, including the access to information not included within the article text, to inspect the claims made in the article, draw independent conclusions, and share the gained information with others. 

\textbf{To probe different aspects of data that are not mentioned in the article text.}
Participants discussed that the information obtained from reading the visualizations allows them to observe aspects of the data that are not mentioned in the news article. For example, P14 expressed their frustration that \textit{``Even though I get a general understanding, I want to know more than what the author of the article wants you to know. What about the Southeast or the Northeast?''} This sentiment was also observed from other participants. P17 noted that \textit{``The article talks about a couple of different examples, whereas the visualization has far more industries. So if you wanted to look at an industry that wasn't highlighted in the article, it is provided in the visualization.''}

\textbf{To inspect the claims and draw conclusions.}
The desire to inspect the claims made within an article was also widely echoed among participants. As P6 commented, \textit{``The article mentions that there were some decreases in jobs in particular areas and it was helpful that I could look at the visualization and see that number to fully understand what was being discussed.''} Moreover, while inspecting visualizations, participants had the opportunity to confirm the article's claims or build upon the information given within the article. For example, P17 shared that \textit{``Even though I previously read that the trend was increasing gradually overall, after looking at the table, it seems to show that while it is increasing overall, there are some points where the trend flattens out.''} P17 also noted that \textit{``If the claims were to be not honest, I would look at the data.''} These sentiments about authors' bias and misinformation were observed by other participants. P2 shares that \textit{``I am skeptical of any statistic that comes out of nowhere.''} When no data is provided, the participants expressed their inability to do inspect the claim and draw their own conclusions. P9 went further and asserted that their goal to read visualizations was to \textit{``obtain the data''} to draw their own version of conclusions.

\textbf{To share with others.}
Participants expressed that they would like to share the information gained from visualizations with others. P14 noted that \textit{``I would like the important information in a visualization because I may want to share it with someone sighted and explain what the visualization is conveying.''} Within the aspect of sharing the gained information from a visualization, P14 mentioned that \textit{``I would like to know the trends if I was to share it with someone else.''} 

\subsubsection{Constructing Mental Model of Visualizations}
Participants expressed the desire to construct a mental model of a visualization that is similar to what sighted people would view in order to visually interpret the visualizations. Creating a mental image seemed to enhance participants' understanding of visualizations. As mentioned by P9, \textit{``I want to make sure I understand the visualization and where the data points are, by trying to visualize it.''} Similarly, P1 noted that \textit{``I imagine the visualization because I like to understand each piece of information.''} While constructing a mental picture, many participants had follow-up questions regarding the visualization and its details. For example, after having a follow-up discussion regarding the visualization, P10 shared that \textit{``The additional information helped knowing how this graph is laid out and I can visualize what is being described.''} Other participants also stated that having supporting details guided them to construct a better landscape of the visualizations. As demonstrated by P18, \textit{``I was drawing with my hand to try to visualize what the chart might look like based on the given alternative text.''} P12 shared that \textit{``If I had more information about the visualization, it would have helped me to mentally visualize what was depicted.''}

Overall, participants expressed the desire to know enough visual details about the visualization to imagine them. Many participants connected this aspect to the knowledge that a sighted person would have, such as P6, who noted that \textit{``I want to get the same information that a sighted person is getting when looking at a visualization.''} Additionally, P21 shared, \textit{``I want to imitate the way that sighted people would skim the visualization. I would want to see it the way that a sighted person would think about it from their phenomenological experience. I am visualizing what I am missing that would be a normal thought in a sighted person’s head.''} We observed that creating mental visualizations allowed participants to \textit{``think like a sighted person when I am reading an alternative text,''} as explained by P15. Notably, many participants expressed their frustration with the limitations of creating a mental picture of the visualizations. In P14's words, \textit{``It is difficult to visualize the whole graph in my head.''} This sentiment indicates that, although participants attempted to create mental images, constructing and keeping the mental images for long periods is difficult, likely due to the sequential (and non-interactive) nature of audio delivery.

\subsubsection{Information Needs for Alt-texts}
We learned the participants' expectations and needs regarding what descriptive information to include in visualization alt texts.

\textbf{Chart type.}
11 out of 22 participants echoed that providing the type of chart is helpful, especially at the beginning of the alt text. Since they learned charts at school, stating the chart type can prepare them to fill in the missing information specific to each chart type. As P10 mentioned, \textit{``I think the type of chart is helpful. Is this a pie chart or bar chart, line graph, etc. I think that's a good start to guide a person to think of and imagine [the visualization].''}

Another benefit of starting by stating the chart type is to indicate the beginning of an alt text for the visualization. For example, P11 shared \textit{``So [I would give] priority to the type of charts. Just to give you an indication where the chart is. Especially [because] some screen readers only mention 'graphics' at the end when they read the image.''}  

\textbf{Axis, range, \& ticks.}
Information about the axis in a visualization was often highlighted as necessary to understand visualizations and participants often noticed whenever this information was missing (e.g., with brief descriptions).
For example, when P17 was prompted to improve the brief description style of alt texts, they stated: \textit{``What's on the X-axis, what's on the Y-axis. That is for sure.''} P5 also said, \textit{``I didn't hear the X and Y-axis. So, that's a little harder for me to follow.''} The range of values in each axis also communicates boundary statistics of the data set depicted by the visualization, which helps form a more concrete picture. As P7 stated, \textit{``It [range] gives me the boundaries in the framework of what the chart is telling me.''} 

We did not observe evidence that participants wish to know tick information. When we mentioned the axes' increments in the detailed description, no participants pointed out this information to be helpful. Also, while examining the visualization with brief descriptions, insights or tables, which do not contain tick information, no participants noticed their absence. 

\textbf{Data trends.}
Another important component of alt text for visualizations, as highlighted by 14 out of 22 participants, is the description of data trends. Compared to alt texts for images, communicating visualizations have a purpose, often encapsulated by the visible data trends. As P3 shared, \textit{``There's usually a main point that they're trying to emphasize, like the trends in a certain data set. So I think that, for charts, unlike images, it's important for the alternative text to describe the trends that are being demonstrated.''} Participants expressed that authors should describe data trends in visualization alt texts. For example, P1 wanted a summary of the data trends when that information was missing from the alt text \textit{``'Seven-day average is overall increasing [in stimulus V1].' That's kind of the summary I'd put on the map [in stimulus V2].''} P9 mentioned that \textit{``[alt text should include] a visual description of what is being described and some relevant pieces of data, for example, trends or peaks.''} P10 suggested including a ``snapshot'' of the visualization in alt texts: \textit{``A snapshot you can just take a look, and say: okay, that's the trend.''}

In describing data trends, referencing data is essential to provide a concrete picture of the visualization instead of simply describing the pattern. For example, while examining stimulus V4, P1 mentioned, \textit{``[In addition to acknowledging the performance gap, I'd like to know] how wide the gap is between boys and girls. For example, this chart shows the difference of 20\%.''} Specifying the data range can further help comprehend the data trend, as noted by P4: \textit{``[I would describe it as an] upward trend increasing right now, or in the past couple of weeks.''} When depicting the trend, referencing visual attributes such as color to describe trends would not be ideal as people may not readily make sense of them. For example, when we mentioned \textit{``There is a cluster of orange circles positioned relatively top of the chart''} as a part of describing V4, P17 shared that \textit{``[I would prefer it to] just say English and math instead of being the orange and blue circles.''} 

Beyond the trends, 4 out of 22 participants also wanted to be aware of visible outliers, spikes, or dips in the data. For example, P9 shared, \textit{``To be able to visualize it, I would definitely add more of those dates, in particular the spikes or [dates] that have zero [cases].''} Similarly, when prompted for their ideal alt text, P2 suggested, \textit{``I would say a line graph showing the trends with spikes on these certain dates.''} P1 wondered, \textit{``There are some trends from one to three million and it's going up, but are there also some spikes or is it going up overall?''}

\textbf{Colors.}
Participants' preferences toward information regarding colors were varied. Some participants wanted to know, while others did not. Learning about the colors seemed to be a personal preference. Among those who favored knowing the colors, they were especially curious when the color encoded data and when contextualizing the color scale was not challenging. 

To justify their preference, participants often stated that color helps them picture the visualization. It also helps them understand how the visualization uses colors to represent the data. For example, P16 said that \textit{``I personally like that the different colors for the positive and negative changes are mentioned. I just think it levels the playing field a bit.''} P3 shared, \textit{``Those details, like colors, aren't necessarily needed, but I think it helps to beef up my understanding of what the chart looks like.''} On several occasions, participants wanted color information only when it encoded data. For example, when P2 examined V2, she mentioned, \textit{``I don't really care about the colors [green bar]; I just care about how many vaccines are being distributed.''} However, she stated while reading V1, \textit{``With the rest of the information depicted in the table, all you needed [in the alt text] was a color key.''}

The communication of color was especially frowned upon when it encoded a continuous value (V2) using a color gradient schema. In this case, participants seemed to prefer direct access to the values instead of contextualizing them through the color information. As stated by P18, \textit{``If the information in the alt texts were to describe the different colors, that would have just been extraneous. It would have been more information than was necessary. It wouldn't have been necessary because the percentages are there.''}

Regardless of their preferences, participants stated that referencing the colors to explain other aspects of data later in the alt text is overwhelming. For example, P17 shared, \textit{``I do remember that orange was English and blue was math and then later the alt texts refer to the color to explain something. So I had to scroll back up and be like, wait, which one was that?''} To address the cognitive load of the color mapping, participants wished for more intuitive colors that would help them remember the underlying encoding. P5 suggested, \textit{``The positive ones were a light blue and the negative ones were a dark blue, but I would use completely different colors. For the negative ones, I would use a different color like red. That would be helpful to remember.''}

\textbf{Data points.}
All participants favored having access to the underlying data of the visualization. As highlighted by the participants, having access to the data can be beneficial in several ways. It enables participants to interact with the data by seeking out patterns by themselves or finding specific data entries that they might be interested in. P7 shared that \textit{``I'm in Michigan so I wanted to see the data, or what color Michigan was. I wouldn't be able to do that [with the given alt texts].''}
Having access to the data also enhances trust in the authors' claims as, if in doubt, the participants would be able to confirm the claims independently. For example, when asked to comment on an alt text describing the data trends, P14 mentioned, \textit{``I don't mind [the data trends], but I would want to see it for myself anyway.''}

When given a table, participants overwhelmingly preferred the table formatted by HTML table elements (i.e., th, td, tr tags), since they are easy to navigate using the arrow keys with a screen reader. We observed that all participants who were given the HTML formatted table were able to fluently navigate the table without having any trouble. P20 mentioned, \textit{``I think this [HTML formatted] table was really effective. I think the table in some ways was more effective than the other things [textual information].''} P6 echoed the importance of the formatted table: \textit{``Having it separate from the alternative text is important because I could more easily look at it with my screen reader.''}

Participants who were given non-formatted tables inside the alt text stated their frustration for not being able to navigate them. P14 shared, \textit{``The problem is that I cannot use my table navigation commands because this is not an actual table element.''} Another disadvantage of having a non-formatted table is that participants are not able to keep track of where they are at. In the case of a formatted table, when moving from a different column, screen readers will read the column name, then the value. P14 noted, \textit{``I need a table, like a real table, because otherwise I have to constantly remember which column I'm at.''} 

Regarding the table contents, participants stated their preference for sorted tables with the sorting criteria explicitly stated, since such tables would be easier to parse. For example, P3 mentioned, \textit{``The graph is organized from the biggest gain down to zero, and then down to the biggest loss. I like that the table is organized in the same way. If it was out of order, that would be confusing.''} P2 added that \textit{``The table caption should indicate how the table is sorted.''} Clearly defined data columns, potentially with the respective units and a brief description of their meaning or how they are computed, were also deemed necessary to fully understanding the data.
P3 stated, \textit{``I don't really understand some of the numbers. [The achievement gaps] are all numbers less than one. I don't know how those numbers are calculated.''}

\subsubsection{Style Needs for Alt-texts}
\textbf{Length.}
The length suggested by the majority of participants ranges from 2 to 8 sentences. For example, P1 shared that \textit{``I would say from three to six or sometimes eight sentences. I don't mind waiting for information. If I want to read it, I read it.''} However, P5 asserted that more than six sentences is too long: \textit{``I would say, maybe between four and six sentences, because once you get beyond six sentences, it gets to be a lot, and it can be overwhelming and confusing sometimes.''} Some participants specifically stated that it depends on the visualization's complexity and how much information is already covered by the article. For example, P8 mentioned, \textit{``That depends on the chart and what's being described. Sometimes it's such a busy chart that you're going to have to write pages of analysis on what's in it, and it depends on what information is provided in the article as well.''} A few participants mentioned they don't mind the length as long as it has all the information they need. As P6 stated, \textit{``I always want things to be more specific if it can be. So I don't mind if there's long alt text.''}

\textbf{Language.}
Most participants prefer plain terms, as exemplified by P2's comment, \textit{``From an accessibility standpoint [I prefer] simpler terms''}, and P13's preference for \textit{``more of a simplified type of language so that it's not so technical, and it makes it easier.''} We learned that participants did not understand some technical terms, such as gradients. As P14 shared, \textit{``I didn't exactly know what they mean by the gradient. Maybe, explain it in simpler terms, like the shade.''}

Also, we learned that participants prefer an objective tone. P16 shared that \textit{``It almost seems like somebody different than the person writing the article should do the alt text.''} P17 mentioned that decorating the data trend with adjectives sounds less objective: \textit{``It only moved by this much or only decreased by that much, even the word 'only', to me feels like a little bit of an opinion.''}

\textbf{Order \& navigation of alt text.} 
As stated above, participants prefer the chart type to be stated first, together with summary of the visualization. Then, several participants declared their preference for additional detailed information and, more importantly, followed by data. For example, P14 specifically mentioned that \textit{``Ideally [I'd like a] brief description and summarization that covers all the vital points, followed by more data.''}

Participants mentioned that this type of structure could also serve different needs of people, as stated by P17, \textit{``I really liked the structure of having a brief description and then a detailed description so that if somebody didn't want to look at all that, and they just wanted the basics, they could get that.''} In a similar vein, P18 wished for a more structured alt text: \textit{``If it had, 'this is the description of this axis, and this is the description of that axis and here's the description of the bars,' that might make it a little easier.''}

Participants mentioned they wished to have some interactivity. Simple interactivity can be a capability to navigate sentence by sentence. Unlike plain text, some screen readers do not support sentence by sentence reading for alt texts. As P15 stated, \textit{``The only thing I hate about the way that alternative text runs is you can't read it sentence by sentence. Depending on what operating system I'm using, I can't. I wish you could just like read it sentence by sentence.''} 

\subsubsection{Summary}
Our findings demonstrate that participants were trying to \textit{visualize} visualizations in their head while listening to alt texts. We also observed that participants' purposes of understanding visualizations were similar to the purposes that sighted people might have while interacting with visualizations. Therefore, alt texts should be designed to support people with visual impairments to mentally visualize the graphics as well as to complete simple visualization tasks. 

Aligned with the guidelines, participants expressed the importance of chart type, axes, data trend described in alt texts. Our findings highlight the necessity of providing a range of axes to bind their imagination. The guidelines recommend that it is unnecessary to mention the visual attributes of the visualizations, such as the color, which was found to have an opposing response from the participants, as several preferred being told the colors being used. Tables were one of the most important aspects that participants desired, as the guidelines outline. Participants depended on the accessible tables to complete many tasks, such as retrieving the data and inferring the underlying patterns. However, the importance of providing details in the table caption, such as how a measure is calculated and how the table is sorted, was lacking in the guidelines. Contrary to the majority guidelines, most participants preferred more than two sentences of alt text to ensure that all of the necessary information is available in alt texts, even if they may skip. Unlike images often auxiliary to the surrounding contents in online news, visualization carries the complementary information to the presented texts, having participants wish to have access to all the information.

Given the findings, the current practice we observed from Phase 2 is not sufficient to satisfy participants' information needs. For example, especially in the CHI collection, lacking data points would prevent participants from extracting any information other than the mere fact that there is a visualization. Also, lacking chart type information would prevent people with visual impairments from imagining anything, even if the alt text contains all other information about visualization.

\section{Discussion}
\subsection{Recommendations for generating alt texts}
Through the study, we observed that visually impaired participants want to use visualizations to accomplish similar goals that sighted people would have. We also observed that participants actively try to construct a mental model of the visualization in their heads while listening to alt texts. 
To support their goals and the construction of an accurate mental picture through alt texts, we suggest formulating alt texts of visualizations by considering the following:

\vspace{-2mm}
\begin{itemize}[leftmargin=*] 
\setlength\itemsep{-0.3mm}
\item  Indicate the start of the alt text by mentioning the \textbf{chart type}.
The chart type serves as a ``template'' that provides people a starting point for constructing a mental picture of the visualization, helping them fill in missing details later. If the type of chart is beyond a commonly used one (e.g., bubble chart), briefly explain how it looks compared to a common type (e.g., bar chart, line chart, scatterplot) if possible. 
\item Communicate the scope of the data by explaining the \textbf{axes} and their \textbf{range}. Describing the increments of ticks may not be necessary.
\item Explain the visible \textbf{data trends} as if you would provide a ``snapshot'' of the visualization. When describing trends, reference the data instead of visual attributes to avoid the unnecessary cognitive effort to remember mappings (e.g., use ``the English cluster is located below zero'' instead of ``the orange cluster is located below zero''). Specify the range of values in which the trend is observed to convey a concrete scene (e.g., use ``it shows an increasing trend between Jan 3rd to Feb 3rd'' instead of ``it shows an increasing trend'').
\item To enrich the mental picture of visualizations, optionally provide a brief description of the \textbf{mapping} between the data and visual attributes (e.g., color, shape) if applicable. 
\item To accommodate the different amounts of information needs, place a brief description of the visualization first (i.e., chart type with a summary of the topic), then describe it in detail (e.g., axes, range, mapping, trend). Possibly, mention ``Details of the visualization are as follows'' after the brief description so that those who do not wish to continue listening can skip. The detailed description can be placed in the alt text, the \texttt{longdesc} attribute \footnote{Although longdesc is deprecated, most screen readers can read it. If a new standard for long descriptions is introduced, we recommend people to use it.} of an image element, or \texttt{<desc>} tag in an SVG element.
\item Place the \textbf{data table} as a hidden HTML element next to the visualization, and mention at the end of the alt text how to access the data table (e.g., ``The data for this chart is available in the table below''). Alternatively, place the data table on a separate page, and add a link to it in the \texttt{longdesc}. Often, visualization designers sort data by a meaningful variable to make the pattern in the visualization more salient. The same principle should apply to the table design. Sorted tables help people recognize the pattern while navigating the table. Tables laid out in the same order as in the visualization also help people envision the depicted chart with less effort. The table should contain a caption explaining its contents, the sorting criteria, and how the various data fields were calculated. The designer can use the well-crafted labels used in the visualization to name the columns. When possible, prioritize the shortest name that conveys a column's meaning since screen readers repeatedly read the column name when moving between columns. If the table contains too many rows, sample the data and indicate so in the caption. Ideally, stratify the samples based on the critical variables.
\item Use plain language instead of visualization-specific language (e.g., shade vs. gradient) to enhance accessibility. Avoid decorating data (e.g., \textit{only} 10\%, \textit{very large} gap) to maintain objectivity. 
\end{itemize}

Since some screen readers do not fully support the \texttt{desc} tag of an \texttt{svg} element (equivalent to the \texttt{alt} attribute of an \texttt{img} element), we recommend designers to consider creating a dummy image element below the visualization to add an alt text in the \texttt{alt} attribute.

\subsection{Design implication beyond simple alt texts}
Given the apparent limitation of textual descriptions, designers could combine other modalities to complement visualization alt text. We envision a scenario where the alt text describes the visualization components first and sonified data trends are played afterward. As P22 shared, \textit{``[First explain the visualization components] then you could play an audio version of the graph. Like, as the daily doses would spike, the sound would go up and down in pitch.''} Another scenario considers linking the screen reader system with a haptic display to provide tactile feedback when a reader encounters visualizations online. As P5 noted, \textit{``Having a way to convert these visualizations into tactile options such as braille along with the audio of the alternative text would help enhance our form of understanding and learning.''}

Visual retrieval operations, which sighted people can perform with visualizations, are not fully enabled for people with visual impairments. For example, readers may want to obtain information from their own state while looking at a choropleth representing the hunger rate or may wish to compare the doses administrated between two dates. We observed that people with visual impairments achieved this goal by examining the data table. However, locating the values of interest by navigating a table element takes a long time and imposes a cognitive burden. Dynamically personalizing the alt text or the table contents (e.g., stating data for the readers state, city, or zip code based on their IP address) can partially lower their burden of locating a specific value. Accurate Q\&A modules (e.g.,~\cite{kim2020answering}) with natural language query capabilities (e.g.,~\cite{narechania2020}) would drastically reduce the burden of retrieving information from visualizations by allowing people with visual impairments to pose the question they want to answer.

While the practice to provide alt text for images has been promoted, the majority of images on the web lack alt texts~\cite{Stangl2020}. In addition to encouraging designers to provide alt texts, the technologies proposed in the visualization community and other fields can also enhance accessibility by automating the alt text generation process. We envision an alt text generation system that automatically detects the type of chart and the underlying data from a rendered image using the models proposed by Savva et al.~\cite{Savva2011} or Poco and Heer~\cite{poco2017} depending on the chart type. For D3 visualizations, the visual components and the data can be extracted by the model proposed by Harper and Agrawala~\cite{harper2014}. Data trends can also be extracted by automated techniques~\cite{cui2019,law2020characterizing}. Then, the system can use a natural language generation pipeline (e.g., ~\cite{bernardi2016,vinyals2015}) to formulate alt text from the extracted components, following the proposed guidelines.

\subsection{Limitation \& future work}
Due to technical issues of supporting accessible SVG for some screen readers~\cite{bugjaws}, we rendered all of the visualizations in our study into images. As a result, all readable elements, such as labels, ticks, and textual annotations, were lost. As a next step, we wish to investigate how we could leverage readable elements in an SVG, in addition to alt text, to enhance the understanding of visualizations for people with visual impairments. 

In this work, we focus on alt texts for several static charts with varying numbers of variables and encodings, specifically in online document reading scenarios. Future work may expand the scope of the investigation to cover more complex types of visualizations (e.g., uncertainty visualization) and interactive visualizations, as well as other contexts like academic settings, etc. Also, we primarily focused on people without remaining vision. Future work should explore strategies for communicating visualizations for people with low vision. 

Finally, the insights gained from our qualitative investigations can also inform quantitative studies to observe how much the recommended alt text composition can help people with visual impairments complete visualization tasks.

\section{Conclusion}
We analyzed the existing guidelines and current practices for constructing text alternatives for visualizations. We also reported findings from an interview study with 22 people with visual impairments. Our investigation provides insights into how people with visual impairments wish to use visualizations and how they construct an image in their head while listening to alt texts. We identified information needs in visualization alt text to enhance the accessibility of visualizations. Visualizations are a powerful tool to communicate data and their use is pervasive in the media. Thus, ensuring visualization accessibility for people with visual impairments is essential for information equality. We hope that our findings will contribute to the overall accessibility of visualizations.

\bibliographystyle{abbrv}

\bibliography{template}
\end{document}